# Chain connectivity and conformational variability of polymers: Clues to an adequate thermodynamic description of their solutions

## II: Composition dependence of Flory-Huggins interaction parameters

Bernhard A. Wolf*

Institut für Physikalische Chemie und Materialwissenschaftliches Forschungszentrum
der Johannes Gutenberg-Universität Mainz, Jakob Welder-Weg 13, D-55099 Mainz, Germany

**Abstract**

In part I of this contribution we have reported how the Flory-Huggins interaction parameter $\chi$ can be modeled as a function of chain length within the composition range of pair interaction between the macromolecules by means of the three parameters $\alpha$, $\zeta$ and $\lambda$. This contribution presents the extension of the approach to arbitrary volume fractions, $\varphi$, of the polymer and its application to published data on $\chi(\varphi)$. The resulting equation reads $\chi = \alpha(1-\nu\varphi)^{-2} - \zeta(\lambda + 2(1-\lambda)\varphi)$ and requires only the additional parameter $\nu$ to incorporate the composition dependence. Its employment to experimental data is very much facilitated by substituting for $\chi_0$ (limiting value for $\varphi \to 0$); furthermore the expression can in good approximation be simplified to $\chi \approx (\chi_0 + \zeta\lambda)(1-\nu\varphi)^{-2} - \zeta\lambda(1+2\varphi)$. This means that only two parameters, $\nu$ and the product of $\zeta$ and $\lambda$, need to be adjusted. This relation is capable of describing all types of composition dependencies reported in the literature, including the hitherto incomprehensible occurrence of pronounced minima in $\chi(\varphi)$. For a given system the evaluation of the chain length dependence of $\chi_0$, reported in part I, and the present evaluation of the composition dependence of $\chi$ yield the same data for the conformational response $\zeta$. Similarly both types of measurements generate the same interdependence between $\zeta$ and $\alpha$. The physical meaning of the different parameters and the reason for the observed correlations are discussed.

*Keywords*: Flory-Huggins interaction parameter, Composition dependence, Temperature dependence, Conformational relaxation

\* Corresponding author: Tel.: +49-6131-392-2491; fax: +49-6131-392-4640
  *E-mail address*: Bernhard.Wolf@Uni-Mainz.de

## List of symbols

| | |
|---|---|
| $A_2$ | second osmotic virial coefficient |
| $a$ | exponent of the Kuhn-Mark-Houwink equation |
| $G$ | Gibbs energy |
| $g$ | integral Flory-Huggins interaction parameter |
| $K$ | factor of the Kuhn-Mark-Houwink relation |
| $m$ | parameter of Eq. (7) |
| $M$ | molar mass |
| $N$ | number of segments |
| $n$ | parameter of Eq. (7) = $-\nu$ |
| $p$ | parameter of Eq. (7) = $-\zeta$ |
| $q$ | parameter of Eq. (7) = $1 - \lambda$ |
| $s$ | surface of a segment |
| $V$ | volume |
| $w$ | weight fraction |
| $\alpha$ | short for $\chi_{o,f-c}$ |
| $\beta$ | = $\chi_{overlap.\ coil}/\chi_{isol.\ coil}$ |
| $\gamma$ | = $(1 - s_2/s_1)$ |
| $\zeta$ | conformational response to dilution = $\beta - 1$ |
| $\Theta$ | theta temperature |
| $\kappa$ | $K_N \rho_2$ |
| $\lambda$ | = $\dfrac{1}{2} + \kappa N^{-(1-a)}$ |
| $\nu$ | = $-n$ |
| $\xi$ | differential Flory-Huggins interaction parameter referring to the polymer |
| $\rho$ | density |
| $\varphi$ | volume fraction |
| $\Phi$ | average volume fraction of segments within the space defined by the polymer coil |
| $\chi$ | differential Flory-Huggins interaction parameter |
| $[\eta]$ | intrinsic viscosity |

*superscript*

| | |
|---|---|
| $-$ | molar quantity |
| $=$ | segment molar quantity |
| o | infinite dilution |

*subscripts*

| | |
|---|---|
| 1 | solvent |
| 2 | polymer |
| f-c | fixed conformation |
| isol. coil | isolated coil |
| N | referring to polymer segments |
| overlap. coil | overlapping coil |

## I. INTRODUCTION

An adequate theoretical description of the thermodynamic behavior of polymer containing systems is still a challenging task, even half a century after the pioneering work in this field[1] and despite very sophisticated new approaches[2]. Two outstanding examples for deficiencies are named explicitly: In the dilute regime current approaches cannot account for an augmentation of the second osmotic virial coefficient $A_2$ (i.e. for a reduction of the Flory-Huggins interaction parameter $\chi$ ) with rising molar mass of the polymer[3,4]. With the dependence of $\chi$ on the volume fraction $\varphi$ of the polymer, the problem concerns the modeling of experimentally observed minima[5] and the rationalization of several other unaccountable dependencies of $\chi$ on $\varphi$.

In the present series of publications we want to check whether it is possible to overcome some of the existing difficulties by accounting for the connectivity of the monomers of macromolecules and their ability to modify the conformation in response to the changes in their molecular surrounding re-sulting from the mixing process. Part I[6] deals with these phenomena in the case of pair interaction between the solute. In part II we want to extend the approach presented there to higher polymer concentration, covering the entire range up to the melt. To that end we briefly recapitulate the main results obtained for the chain length dependence of $\chi_o$ (the limit of the Flory-Huggins interaction parameter for infinite dilution) first.

## II. THEORETICAL CONCEPT

The present approach allows for the fact that some contacts between segments belonging to the same molecule cannot be opened by the addition of any amount of solvent (as a consequence of chain connectivity) and accounts for the ability of flexible macromolecules to change their conformation[7] in response to a variation in the environment (conformational variability). In contrast to the normal procedure the modeling starts from highly dilute polymer solution well below the region within which more than two polymer coils can overlap. The details of that thoughts are presented in part I of this contribution. Here the results are only briefly recalled to the extent required for the generalization of the approach.

### A. Pair interaction

For $\chi_o$, the Flory-Huggins interaction parameter in the limit of pair interactions between the polymer molecules we have obtained the following expression

$$\chi_o = \alpha - \zeta \left( \frac{1}{2} + \kappa N^{-(1-a)} \right) \quad (1)$$

where $\alpha$ measures the effect of the addition of further solvent molecules (dilution at constant composition) keeping the conformation of the chain containing $N$ segments unchanged. The second term accounts for the conformational relaxation of the polymer in response to the insertion of a solvent molecule between two contacting segments (belonging to different molecules). The parameter $\zeta$, called conformational response, becomes zero in the special case of theta solvents. The factor $\kappa$ is given by

$$\kappa = K \rho_2 \left( \frac{\rho_2}{\rho_1} M_1 \right)^a \quad (2)$$

where $K$ and $a$ are the usual Kuhn-Mark-Houwink parameters, relating intrinsic viscosities with molar masses; $\rho_i$ are the densities of the components and $M_1$ is the molar mass of the solvent. Due to the fact that we are presently interested in the composition dependence of $\chi$ for a given polymer with constant chain length, we can treat the expression in the brackets of Eq. (1) as a constant (which is in most cases only slightly larger than 0.5). Setting

$$\left( \frac{1}{2} + \kappa N^{-(1-a)} \right) = \lambda \quad (3)$$

we can rewrite Eq. (1) as

$$\chi_o = \alpha - \zeta \lambda \quad (4)$$

The first term of this relation measures the non-combinatorial Gibbs energy of dilution resulting from the insertion of a solvent molecule between two contacting polymer segments belonging to different chains; the conformation of the two involved polymer molecules and the composition of the system are kept constant. This procedure does not yet establish the chemical equilibrium. In order to reach that state, the conformation of the solute molecules must still adjust to the new situation (conformational relaxation). The corresponding contribution to $\chi_o$ is quantified by the second term of Eq. (4). According to the results of part I[6], $\alpha$ and $\zeta$ assume large positive values in most cases. Although it is obvious that both parameters should contain - as usual - enthalpy and entropy contributions, $\alpha$ ought to be dominated by enthalpy and $\zeta$ by entropy.

## B. Composition dependence

The extension of the relations of the last section, formulated for highly dilute solutions in terms of the usual Flory-Huggins interaction parameter (defined via the chemical potential of the solvent), will be performed in terms of the integral interaction parameter $g$. A direct generalization of the expression obtained for $\chi$ would be considerably less straightforward for mathematical reasons that become obvious from Eq. (5), in which $\varphi$ stands for the volume fraction of the polymer.

$$\chi = g - (1-\varphi)\frac{\partial g}{\partial \varphi} \qquad (5)$$

The analogous relation between the integral parameter $g$ and the differential interaction parameter $\xi$, referring to the polymer, reads

$$\xi = g + \varphi\frac{\partial g}{\partial \varphi} \qquad (6)$$

From the results obtained in part I (Eq. (4)) it is obvious that the corresponding expression $g(\varphi)$ ought to consist of at least two terms, which must of course reproduce the original relation for $\chi_o$ after differentiation (Eq. (5)) and simplification for $\varphi \to 0$. The first summand should depend on composition for no less than two reasons. Firstly, because of the differences in the surfaces of solvent molecules and polymer segments (where the segment is as usual defined in its size by the volume of the solvent). On the basis of model considerations this effect was formulated[8] as a proportionality to $(1 - \gamma \varphi)^{-1}$. The second reason why the first summand may depend on composition results from the fact that the deviation from combinatorial behavior should become smaller as the degree of coil overlap increases. If we assume that the same mathematical relation holds true for both contributions we can write $(1+n\varphi)^{-1}$, where $n$ encompasses both, surface and entropy effects.

The introduction of a composition dependence for the second term is straightforward and based on the obvious fact that the number of segments that can be separated by the addition of a solvent molecule must in the general case increase with polymer concentration. This implies that more than two chains contribute to the effect resulting from dimensional relaxation at higher concentrations. The factor $(1 + q\varphi)$ is introduced to model this feature. The simplest possible relation for $g(\varphi)$ thus reads:

$$g = \frac{m}{1+n\varphi} + p(1+q\varphi) \qquad (7)$$

where $m$, $n$, $p$ and $q$ represent system specific parameters.

In order to control whether the above *Ansatz* assumes the same form as Eq. (4), we calculate $\chi$ by means of eq (5) and obtain

$$\chi = \frac{m(1+n)}{(1+n\varphi)^2} + p(1+2q\varphi - q) \qquad (8)$$

which becomes

$$\chi_0 = m(1+n) + p(1-q) \qquad (9)$$

for infinite dilution.

Comparison of Eqs. (9) and (4) yields the following relations

$$m(1+n) = \alpha \qquad (10)$$

$$p = -\zeta \qquad (11)$$

and

$$1 - q = \lambda \qquad (12)$$

Introducing $-v$ instead of n

$$v = -n \qquad (13)$$

to attain positive values for all typical parameters, we can reformulate Eq. (9) as

$$\chi = \frac{\alpha}{(1-v\varphi)^2} - \zeta(\lambda + 2(1-\lambda)\varphi) \qquad (14)$$

It can be easily seen that this expression reduces to Eq. (4) as $\varphi$ approaches zero.

Rewriting the integral interaction parameter and that referring to the chemical potential of the polymer in the new nomenclature, i.e. in terms of experimentally accessible parameters, yields

$$g = \frac{\alpha}{(1-v)(1-v\varphi)} - \zeta(1+(1-\lambda)\varphi) \quad (15)$$

and

$$\xi = \frac{\alpha}{(1-v)(1-v\varphi)^2} - \zeta(1+2(1-\lambda)\varphi) \quad (16)$$

## III. EVALUATION OF EXPERIMENTAL DATA

For the application of Eq. (14) to primary data on $\chi(\varphi)$ it is advisable to reduce the number of parameters, particularly in view of the findings of part I according to which some of them are interdependent. Because of the high accuracy with which $\chi_o$ is normally obtained from light scattering or osmotic experiments, the most obvious improvement is the substitution of this quantity, which reduces the number of parameters to three. The next option for simplification is $\lambda$. There are two reasons for this choice: (i) For the description of the composition dependence of $\chi$ the chain length of the polymer is predetermined so that $\lambda$ becomes a constant. (ii) According to theoretical considerations and experimental findings $\lambda$ varies little. This is so because typical $\kappa$ values range from 0.2 ($a = 0.8$) to 0.6 ($a = 0.5$) and even for a low $N = 100$ the term $\kappa N^{(1-a)}$ lies in the range from 0,08 to 0,06 corresponding to $\lambda$ changes from 0,58 to 0.56. Furthermore $\lambda$ approaches 0.5 rapidly as $N$ increases.

In view of the above-said and the fact that it is often difficult to determine $\zeta$ and $\lambda$ separately from the composition dependence of $\chi$ (in contrast to the situation with $A_2(M)$, where this is possible), we rewrite Eq. (14) as

$$\chi = \frac{\chi_0 + \zeta\lambda}{(1-v\,\varphi)^2} - \zeta\lambda\left(1 + 2\left(\frac{1}{\lambda}-1\right)\varphi\right) \quad (17)$$

and set $\lambda = 0.5$ in the inner bracket, so that we end up with

$$\chi \approx \frac{\chi_0 + \zeta\lambda}{(1-v\,\varphi)^2} - \zeta\lambda(1+2\varphi) \quad (18)$$

This means that only two adjustable parameters are required, namely $\zeta\lambda$ and $v$, to model the composition dependence of $\chi$ according to the present approach.

Due to the large expenditure associated with the measurement of Flory-Huggins interaction parameters as a function of composition, data on that dependence are scarce. Two polymers were chosen to check the validity of the present approach. Poly(vinyl methyl ether)[5] [PVME], because of the uncommon, so far unexplained $\chi(\varphi)$ passing a pronounced minimum, and polystyrene [PS] for the reason that it is the only polymer for which information on its behavior in different solvents is available[9].

### A. Poly(vinyl methyl ether)

Figure 1 shows the original data $\chi(\varphi)$ for the system cyclohexane/poly(vinyl methyl ether) [CH/PVME] together with the curves resulting from an adjustment of $v$ and $\zeta\lambda$ according to Eq. (18). These graphs demonstrate that the experimental results can be well modeled by these two parameters, which are collected in Table 1. Their physical relevance and their variation with temperature is discussed in section IV

.

**Table 1:**

Parameters obtained from the modeling of the concentration dependence of the Flory-Huggins interaction parameter for solutions of PVME of different molar mass in CH at the indicated temperatures according to Eq. (18). The molar masses (kg/mole) of the polymers are stated together with their acronyms.

|  | T/°C | $\chi_o$ | $\alpha$ | $\zeta\lambda$ | $\nu$ |
|---|---|---|---|---|---|
| CH/PVME 28 | 35 | 0.525 | 1.599 | 1.074 | 0.398 |
|  | 45 | 0.510 | 1.561 | 1.051 | 0.377 |
|  | 55 | 0.470 | 1.521 | 1.051 | 0.377 |
|  | 65 | 0.432 | 1.735 | 1.303 | 0.380 |
| CH/PVME 51 | 35 | 0.525 | 1.531 | 1.006 | 0.398 |
|  | 45 | 0.520 | 1.538 | 1.018 | 0.380 |
|  | 55 | 0.480 | 1.547 | 1.067 | 0.383 |
|  | 65 | 0.437 | 1.621 | 1.184 | 0.377 |
| CH/PVME 81 | 35 | 0.535 | 1.232 | 0.697 | 0.362 |
|  | 45 | 0.519 | 1.232 | 0.713 | 0.354 |
|  | 55 | 0.482 | 1.239 | 0.757 | 0.342 |
|  | 65 | 0.446 | 1.256 | 0.810 | 0.339 |

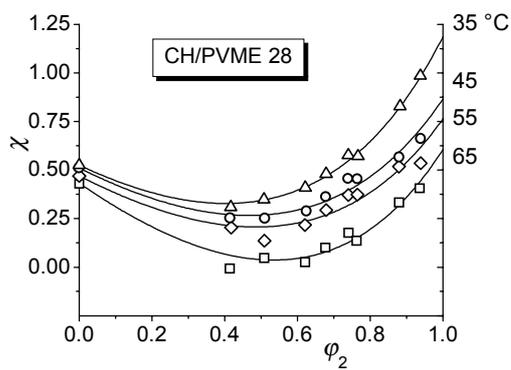

Fig. 1, part a

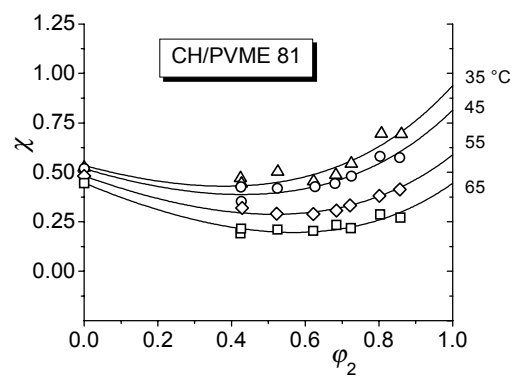

Fig. 1, part b

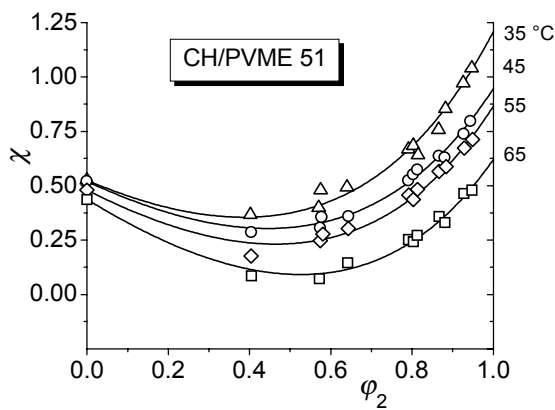

Fig. 1, part c

**Figure 1**. Evaluation of the Flory-Huggins interaction parameters $\chi$ as a function of $\varphi$, the volume fraction of the polymer, according to Eq. (18) for solutions of PVME in cyclohexane[5]. The molar masses (kg/mole) of PVME and the temperatures are indicated in the graphs.

## B. Polystyrene

The probably best studied solvent for this polymer is the theta solvent cyclohexane (CH). The evaluation of the data by analogy to the for PVME is presented in Figure 2. The resulting parameters are given in Table 2.

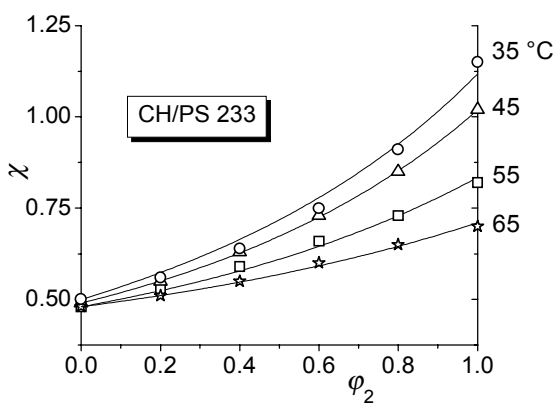

**Figure 2**: Measured[5,9] and calculated (Eq. (18)) concentration dependence of $\chi$ for the system CH/PS 233 kg/mol for the indicated temperatures. The molar mass (kg/mole) of the polymer is given in its acronym.

Other solvents for which information can be found in the literature are very similar in their chemical nature to the monomeric unit of PS; these are benzene (BZ), toluene (TL), and ethyl benzene (EB). Solutions of PS in either of these three solvents or in trichloromethane (TCM) constitute rare examples for systems exhibiting a linear composition dependence of $\chi$.

Figure 3 displays the results for BZ and TL at room temperature and Figure 4 for TCM at 50 °C. All three systems are very exceptional due to the fact that $\chi$ *decreases* considerably with rising polymer concentration. Figure 5 shows the data for EB close to its boiling point and is indicative for solutions for which $\chi$ increases linearly, in contrast to most cases where the increase is considerably more pronounced. The parameters obtained for the different systems are collected in Table 2.

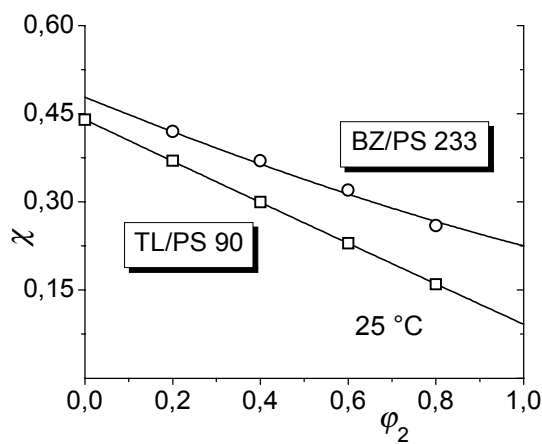

**Figure 3**: Measured and calculated (Eq. (18)) concentration dependence of $\chi$ for the systems benzene/PS[9,10] and toluene/PS[9-11] at 25 °C. The molar masses (kg/mole) of the polymers are stated together with their acronyms.

**Table 2:**

Parameters obtained from the modeling of the concentration dependence of the Flory-Huggins interaction parameter for solutions of PS in different solvents at the indicated temperatures according to Eq. (18). The molar masses (kg/mole) of the polymers are stated together with their acronyms.

|  | T/°C | $\chi_o$ | $\alpha$ | $\zeta \lambda$ | $\nu$ |
|---|---|---|---|---|---|
| CH/PS 233 | 35 | 0.500 | 0.500 | 0.000 | 0.334 |
|  | 45 | 0.490 | 0.523 | 0.063 | 0.314 |
|  | 55 | 0.480 | 0.520 | 0.075 | 0.258 |
|  | 65 | 0.480 | 0.523 | 0.081 | 0.206 |
| BZ/PS 234 | 25 | 0.473* | 0.600 | 0.130 | 0.002 |
| TL/PS 90 | 25 | 0.440 | 0.598 | 0.158 | - 0.063 |
| TCM/PS 290 | 50 | 0.560* | 0.961 | 0.401 | 0.121 |
| EB/PS 275 | 130 | 0.180* | 0.297 | - 0.117 | 0.136** |

\*    Extrapolated

\*\*   Due to the linear dependence and comparatively few experimental data this result should not be overvalued; even negative $\nu$ cannot be excluded. However, $\zeta \lambda$ remains negative in all cases and $\chi_o$ does practically not change.

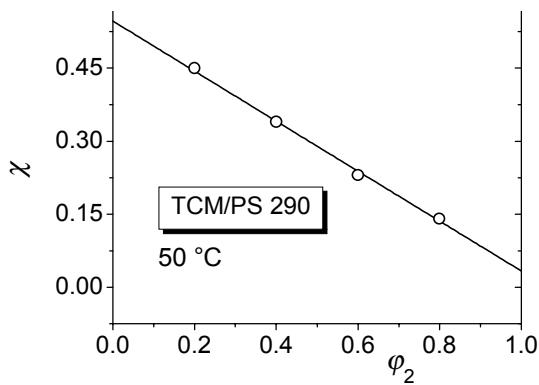

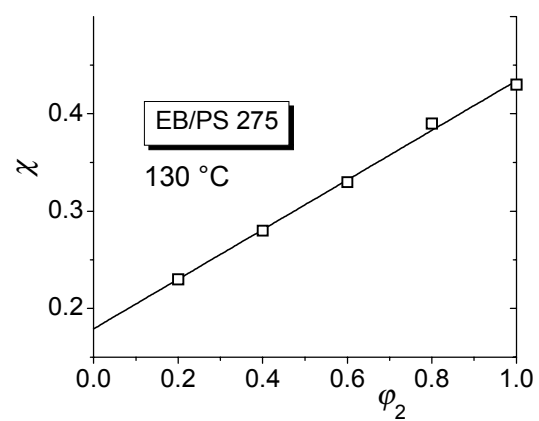

**Figure 4**: Measured and calculated (Eq. (18)) concentration dependence of $\chi$ for the systems trichloromethane/PS[9,12] at 50 °C. The molar mass (kg/mole) of the polymer is stated together with its acronym.

**Figure 5**: Measured and calculated (Eq. (18)) concentration dependence of $\chi$ for the systems ethyl benzene/PS[9,13] at 130 °C. The molar mass (kg/mole) of the polymer is stated together with its acronym.

## IV. DISCUSSION

The results presented in the last section demonstrate the high versatility of the approach based on chain connectivity and conformational variability. Above all it offers an understanding for the existence of minima in $\chi(\varphi)$, as observed with the system CH/PVME, and of $\partial\chi/\partial\varphi < 0$ over the entire composition range, typical for some solutions of PS. Furthermore, it can explain the experimental observation that solvents for which $\chi_o \geq 0.5$ (CH/PDMS at low temperatures and TCM/PS at 50 °C) do not necessarily constitute marginal solvents (theta solvents) or precipitants.

The reason for the occurrence of minima in the composition dependence of $\chi$ results from the concurrence of large $\zeta$ values and large $\nu$ values. Under these condition the second term of Eq. (14) dominates at low polymer concentrations. The favorable effects of conformational relaxation, measured by $\zeta$, dominate and lead to negative $\partial\chi/\partial\varphi$ values. On the other end of the composition range the first term becomes determining, because of the division of $\alpha$ by a small number, and $\partial\chi/\partial\varphi$ assumes positive values.

A different combination of parameters, namely large $\zeta$ values in conjunction with negligibly small $\nu$ values, leads to the practically linear decrease of $\chi$ with $\varphi$, as formulated in Eq. (14). This behavior is for instance observed for solutions of PS in either benzene, toluene or trichloromethane.

In order to demonstrate the absence of phase separation despite $\chi_o$ values larger than 0.5, we translate the information concerning the thermodynamic particularities of the system CH/PVME 81 at 35 °C, so far presented in terms of $\chi$ (by means of Eq. (14) or the simplified version Eq.(18)), into the integral interaction parameters $g$ (Eq. (15)) and, for the sake of completeness, also into $\xi$, the differential interaction parameter for the polymer (Eq. (16)). First of all, the results of this evaluation, displayed in Figure 6, make obvious that the occurrence of a minimum in the composition dependence is in the present case restricted to $\chi$. The other two parameters, $g$ and $\xi$, increase monotonously with rising polymer concentration. The fact that $\chi$ becomes almost identical with $g$ in the limit of $\varphi \to 0$ is only accidental and a distinctiveness of the present system.

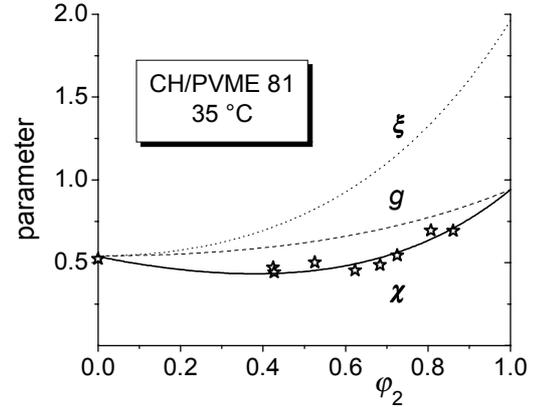

**Figure 6**: Concentration dependence of the Flory-Huggins interaction parameter $\chi$ for the system CH/PVME 81 at 35 °C (experimental data[5] and modeling according to Eq. (18)) and of the corresponding calculated integral interaction parameters $g$ and $\xi$ (Eqs. (15) and (16)). The molar mass (kg/mole) of the polymer is stated together with its acronym.

Knowing $g(\varphi)$ for the two examples exhibiting $\chi_o > 0.5$ (CH/PVME+TCM/PS) we can easily check the existence or absence of a two phase region, by investigating the composition dependence of the reduced Gibbs energy of mixing per one mole of segments according to

$$\frac{\Delta\overline{\overline{G}}}{RT} = (1-\varphi)ln(1-\varphi) + \frac{\varphi}{N}ln\varphi + g(1-\varphi)\varphi \qquad (19)$$

In case of phase separation this function must exhibit points of inflection (spinodal conditions, indicative for demixing). Such points are, however, absent, as can be directly seen from the curves shown in Figure 7, despite $\chi_o > 0.5$ because of the particular composition dependence of $g$. Neither the CH/PVME 81 at 35 °C nor TCM/PS 260 at 50 °C phase separates. Even for infinitely long chains there is no indication of demixing. According to this result the stipulation that $\zeta$ be zero under theta conditions seems to be a more stringent criterion than $A_2 = 0$ (equivalent to $\chi_o = 0.5$).

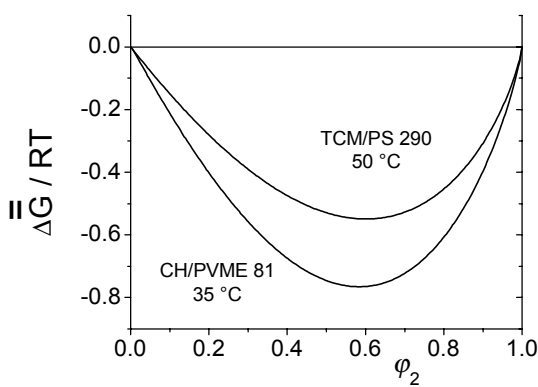

**Figure 7**: Reduced Gibbs energy of mixing for one mole of segments as function of composition for the systems CH/PVME 81 at 35 °C and TCM/PS 260 at 50 °C, demonstrating the absence of a two phase region despite $\chi_o > 0.5$. The molar masses (kg/mole) of the polymers are stated together with their acronyms.

The conformational variability of polymers together with chain connectivity constitutes the pivotal point of the present approach. The consistency of information on the conformational response $\zeta$, as obtained in two fundamentally different manners, therefore represents an important criterion to check whether the model is adequate to describe reality. One source for $\zeta$ is the chain length dependence of the second osmotic virial coefficient (as demonstrated in part I), the other consists in the composition dependence of the Flory-Huggins interaction parameter (this part). It is self-evident that the corresponding data must not contradict each other. Unfortunately the number of systems for which the actual situation can be checked is limited. For three of the PS containing systems studied here, data on $A_2(M)$ were already assessed with respect to conformational relaxation in the literature[4]. In that work the evaluation was performed (Fig. 10 of reference [4]) in terms of $M$, instead of $N$, the information reported on $(\beta - 1)$ must therefore be multiplied by the factor $(\rho_2 M_1/\rho_1)^{-(1-a)}$.

If the directly accessible product $\zeta \lambda$ of the analysis of $\chi(\varphi)$ is split into the individual factors by a reasonable estimation of $\lambda$ according to Eq. (3), one obtains the following $\zeta$ values from the chain length dependence of $A_2$ on one hand and from the compositions dependence of $\chi$, on the other. For CH at 50 °C these values are 0.07 *vers*. 0.07, for BZ at 30 °C 0.42 or 0.30 (according to two different literature sources on $A_2(M)$) *vers*. 0.26 and for TL at 30 °C 0.29 *vers*. 0.30. For two of the systems the agreement is better than expected in view of the accumulating experimental uncertainties with the measurement of $A_2(M)$ and of $\chi(\varphi)$. With BZ/PS, for which the discrepancies are considerable, one must keep in mind that information on $\chi_o$ and very high polymer concentrations is lacking and that the $\zeta$ values obtained for that system are therefore necessarily rather inaccurate.

One of the most striking features of the present approach is the interrelation between $\zeta$ and $\alpha$ reported in part I. Here we cannot reliably separate $\zeta$ from the product $\zeta \lambda$, as discussed earlier. Because of the minor variation of $\lambda$ as compared with $\zeta$ we plot in Figure 8 the product as a function of $\alpha$. In view of the agreement of the situation described in the last section it is not surprising that the present data again fall on a common line. These relations state that the two terms in Eq. (4) are not independent of each other. A deterioration of solvent quality due to $\alpha$ corresponds to an improvement resulting from $\zeta \lambda$. The reasons for this behavior are discussed in the next section.

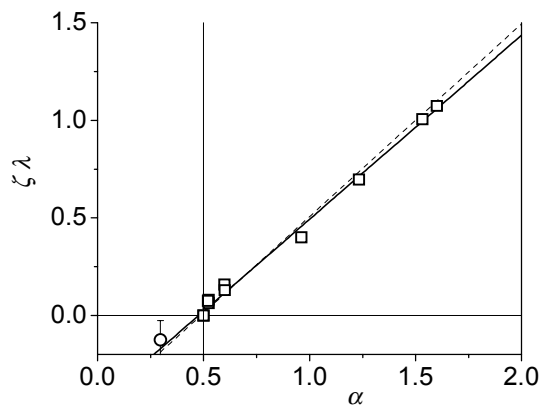

**Figure 8**: Interrelation (open squares and full line) between $\zeta \lambda$ and $\alpha$ for the systems evaluated here (cf. Tables). The data point for the EB/PS (open circle) is rather uncertain due to the almost linear dependence of $\chi$ on $\varphi$ and the lacking information concerning $\chi_o$. Also shown (broken line) is the analogous dependence obtained from the chain length dependence of the second osmotic virial coefficient (part I); for the transformation of the results reported there, $\lambda$ was set 0.52 (this is a reasonable average value for the molar masses of interest).

### A. The parameters $\alpha$ and $\zeta \lambda$

Let us first recall the order of magnitudes of the parameters. The $\alpha$ and $\zeta \lambda$ values obtained from the composition dependence of $\chi$ span the range from $\alpha = 0.297$ and $\zeta \lambda = -0.117$ for EB/PS 275 at 130 °C to $\alpha = 1.735$ and $\zeta \lambda = 1.303$ at 65 °C for CH/PVME 28. These intervals are about half as large as that obtained from the dependence of the second osmotic virial coefficients on the chain length of the polymer reported in part I for different systems.

According to the present concept the first step of dilution consists in the insertion of a solvent molecule between two contacting segments *belonging to different solute molecules*. The formation of new contacts is naturally associated with a certain heat of dilution, but at the same time it should also have considerable entropic consequences. The reason is that the opening of intersegmental contacts of the present kind disengages the two involved polymer molecules from their partners and thus makes more of the total volume of the system individually, i.e. separately, accessible to each of them. This increases the number of ways how to arrange the molecules and consequently contributes to higher entropies of dilution. Whether enthalpy or entropy dominates should become obvious from the temperature dependence of $\alpha$. Figure 9 shows that dependence for the two systems on which we have access to this information.

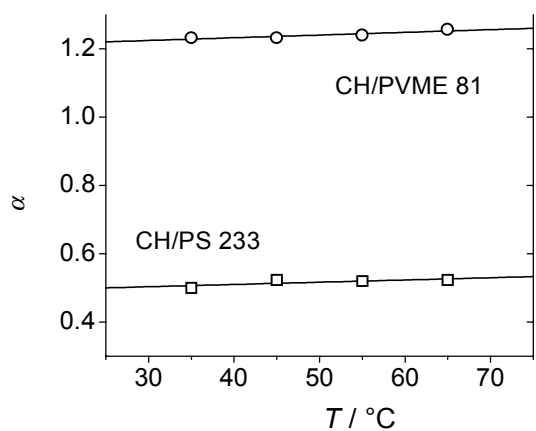

**Figure 9**: Temperature dependence of $\alpha$ for the indicated systems. The molar masses (kg/mole) of the polymers are stated together with their acronyms.

Let us now ask how $\alpha$ varies with the thermodynamic quality of the solvent. For unfavorable solvents the probability to find segments of further macromolecules within the realm of a given polymer coil is higher and for favorable solvents lower that expected from purely statistical considerations due to the establishment of quasi-chemical equilibria. For this reason we can expect smaller $\alpha$ values for bad solvents than for good solvents. This seemingly conflicting statement can be rationalized if one keeps in mind that the liberation of two molecules from a cluster is entropically more efficient than their release from a state in which the solute molecules are already kept apart as much as possible by favorable interaction between the solvent molecules and the polymer segments. So, in conclusion we can state that $\alpha$ increases with rising solvent quality.

The second step of dilution consists in the conformational relaxation, i.e. in the adjustment of the polymer segments contained in the participating coils to the new environment created in the first step. This response to the removal of segments belonging to different molecules from the overlap zone of coils comprises enthalpy and entropy contributions again. Depending on the particular thermodynamic situation the re-establishment of the quasi-chemical equilibrium will prove more or less favorable. For poor solvents the reaction consists in the formation of new contacts between the remaining segments (coil shrinkage upon dilution) whereas more contacts between solvent molecules and polymer segments are being formed (coil expansion) for good solvents. The first case is energetically and entropically considerably less favorable (smaller $\zeta \lambda$ values) as compared with the second case. The temperature dependence of $\zeta \lambda$ (neglecting the contributions of $\lambda$ as discussed later) is shown in Figure 10.

The bottom line of the present considerations concerning the influences of solvent power on $\alpha$ and $\zeta \lambda$ reads that the effect is qualitatively the same for both quantities. The better the solvent for a given polymer becomes the larger the values of the parameters $\alpha$ and $\zeta \lambda$ turn out to be. The present thermodynamic considerations make clear why there should exist an interrelation between these parameters of the type shown in Figure 8 for solutions of a given polymer in different solvents. From the fact that all data fall within experimental error on the same line, irrespective of the chemical nature of the polymer, we can conclude that the differences between the vinyl polymers considered here are only little.

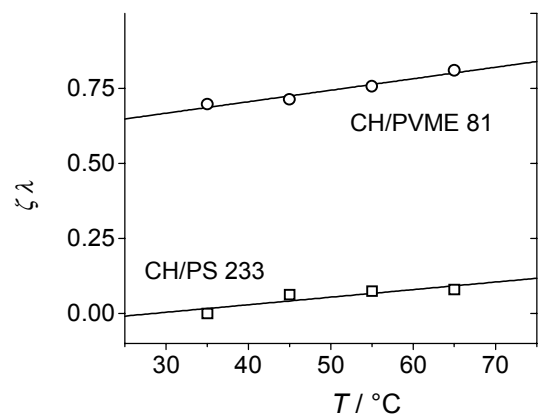

**Figure 10**: Temperature dependence of $\zeta \lambda$ for the indicated systems. The molar masses (kg/mole) of the polymers are stated together with their acronyms.

## B. The parameters $\nu$ and $\lambda$

The $\nu$ values that are obtained from the measured composition dependence of the interaction parameter range from – 0.063 for TL/PS 90 at 25 °C to 0.398 for CH/PVME 28 at 35 °C. The corresponding $\lambda$ values are only accessible with considerable uncertainty from the evaluation of $\chi (\varphi)$ unless the polymer molar mass is varied suitably. On the other hand this parameter is according to its definition (cf. Eq. (3)) even for low molecular weight polymers only slightly larger than 0.5 and varies little with temperature. This means that we can for the present purposes set $\lambda = 0.5$ without substantial loss. This is of course also the reason why we have adjusted $\lambda$ and $\zeta$ conjointly.

We are now going to look after the physical meaning of $\nu$. The term $(1 + n \varphi) = (1 - \nu \varphi)$ in the denominator of Eq. (7) is in its mathematical form identical with an expression introduced long ago to account for the differences in the surfaces of the polymer segments and the solvent molecules[8]. The parameter corresponding to $\nu$ of that approach was called $\gamma$ and introduced as

$$\gamma = 1 - \frac{s_2}{s_1} \qquad (20)$$

where $s$ stands for the surfaces or coordination numbers of the polymer segment (index 2) and the solvent (index 1). Here we refrain from using the same nomenclature because some of the observations can hardly be reconciled with the molecular picture underlying $\gamma$, despite the fact that others match reasonably. The disagreement applies above all the temperature dependence of $\nu$ depicted in Figure 11 for two systems. Even if one interprets the changes in the free volume resulting from an augmentation of temperature as an increase in the surface of the solvent molecules, one cannot rationalize the pronounced reduction of $\nu$. From the present results one may tentatively conclude that the temperature dependence of $\nu$ reflects the sign and the magnitude of the heat of dilution.

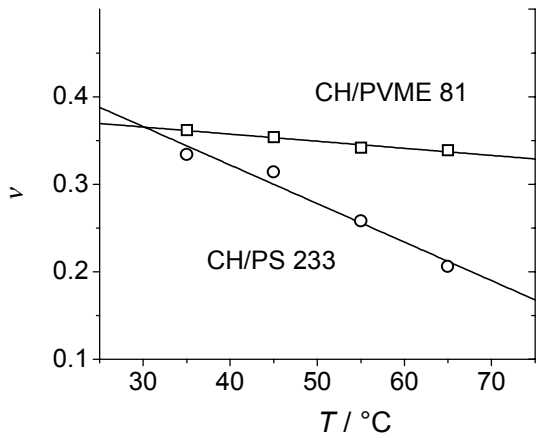

**Figure 11**: Temperature dependence of $\nu$ for the indicated systems. The molar masses (kg/mole) of the polymers are stated together with their acronyms.

On the other hand, for some of the systems the $\nu$ values determined for room temperature lie on the order of magnitude expected for $\gamma$. Solutions of PS in BZ or TL, for which $\nu$ is close to zero, are examples. Even for chemically very similar components, however, some discrepancies are noted, like with the system EB/PS at 130 °C where $\nu$ is much higher than anticipated.

## V. CONCLUSIONS

Accounting explicitly for the chain connectivity and conformational variability of linear macromolecules, relations were established to describe the dependence of the Flory-Huggins interaction parameter on chain length and polymer concentration. In contrast to the normal procedure the present treatment starts from highly dilute solutions with the modeling of hitherto not understood molecular weight dependencies of the second osmotic virial coefficients (part I). The relations obtained in this context are then generalized to comprise the entire composition range (part II). The central parameter introduced in this context is $\zeta$, the conformational response to dilution, which only becomes zero under particular conditions that are normally prevailing in theta systems. All (so far limited) experimental findings can be smoothly rationalized by the new approach. The strongest case in favor of the validity of the established model consists in the fact that the $\zeta$ values obtained for a given system and temperature form the molecular weight dependence of $A_2$ matches well with the corresponding information extracted from the composition dependence of $\chi$ for a given constant molar mass.

The discussion of the detailed thermodynamic meaning of the central parameters is just commencing. One of the outcomes that demands further thinking concerns the definition of theta conditions. As it looks the requirement of $\zeta = 0$ is more general and stringent than $\chi_o = 0.5$. Another important finding consists in the interrelation between $\zeta$ (or $\zeta\lambda$) and $\alpha$ (measuring the effect of dilution at constant conformation of the polymer chain). This feature should turn out helpful for a deeper understanding of the phenomena and for a prediction of $\chi(\varphi)$. A broadening of the experimental basis would be required for a more detailed analysis of the influences of temperature and chain length. Some aspects have not yet been even touched, like the effects of free volume on the different parameters or the possibilities to use the solubility parameter theory or adequate group contribution methods to gain access to them.

**ACKNOWLEDGEMENTS**

The author is grateful to the Deutsche Forschungsgemeinschaft for financial aid.